\documentclass[prb,twocolumn,aps,floatfix,groupedaddress,superscriptaddress,showpacs]{revtex4}
\usepackage[dvips]{graphicx}
\usepackage{rotating}
\begin{document}
\title{Paramagnetic Spin Correlations in CaFe$_2$As$_2$ Single Crystals}
\date{\today}

\author{S.O. Diallo}
\affiliation{Ames Laboratory and Department of Physics and Astronomy, Iowa State University, Ames, IA 50011 USA}
\author{D.K. Pratt}
\affiliation{Ames Laboratory and Department of Physics and Astronomy, Iowa State University, Ames, IA 50011 USA}
\author{R.M. Fernandes}
\affiliation{Ames Laboratory and Department of Physics and Astronomy, Iowa State University, Ames, IA 50011 USA}
\author{W. Tian}
\affiliation{Ames Laboratory and Department of Physics and Astronomy, Iowa State University, Ames, IA 50011 USA}
\author{J.L. Zarestky}
\affiliation{Ames Laboratory and Department of Physics and Astronomy, Iowa State University, Ames, IA 50011 USA}
\author{M. Lumsden}
\affiliation{Oak Ridge National Laboratory, Oak Ridge, TN 37831 USA}
\author{T.G. Perring}
\affiliation{ISIS Facility, Rutherford Appleton Laboratory, Chilton, Didcot, Oxon OX11 OQX, United Kingdom}
\author{C.L. Broholm}
\affiliation{Department of Physics and Astronomy, Johns Hopkins University, Baltimore, MD 21218 USA}
\author{N. Ni}
\affiliation{Ames Laboratory and Department of Physics and Astronomy, Iowa State University, Ames, IA 50011 USA}
\author{S.L. Bud'ko}
\affiliation{Ames Laboratory and Department of Physics and Astronomy, Iowa State University, Ames, IA 50011 USA}
\author{P.C. Canfield}
\affiliation{Ames Laboratory and Department of Physics and Astronomy, Iowa State University, Ames, IA 50011 USA}
\author{H.-F. Li}
\affiliation{Ames Laboratory and Department of Physics and Astronomy, Iowa State University, Ames, IA 50011 USA}
\author{D. Vaknin}
\affiliation{Ames Laboratory and Department of Physics and Astronomy, Iowa State University, Ames, IA 50011 USA}
\author{A. Kreyssig}
\affiliation{Ames Laboratory and Department of Physics and Astronomy, Iowa State University, Ames, IA 50011 USA}
\author{A.I. Goldman}
\affiliation{Ames Laboratory and Department of Physics and Astronomy, Iowa State University, Ames, IA 50011 USA}
\author{R.J. McQueeney}
\affiliation{Ames Laboratory and Department of Physics and Astronomy, Iowa State University, Ames, IA 50011 USA}

\pacs{74.70.-b,75.30.Et,78.70.Nx}

\begin{abstract}
Magnetic correlations in the paramagnetic phase of CaFe$_{2}$As$_{2}$ ($T_{N}=172$ K) have been examined by means of inelastic neutron scattering from $180$ K ($\sim 1.05T_N$) up to 300 K ($1.8T_{N}$). Despite the first-order nature of the magnetic ordering, strong but short-ranged antiferromagnetic (AFM) correlations are clearly observed. These correlations, which consist of quasi-elastic scattering centered at the wavevector $\mathbf{Q}_{\mathrm{AFM}}$ of the low-temperature AFM structure, are observed up to the highest measured temperature of 300 K and at high energy transfer ($\hbar\omega>$ 60 meV). The $L$ dependence of the scattering implies rather weak interlayer coupling in the tetragonal $c$-direction corresponding to nearly two-dimensional fluctuations in the $(ab)$ plane.  The spin correlation lengths within the Fe layer are found to be anisotropic, consistent with underlying fluctuations of the AFM stripe structure. Similar to the cobalt doped superconducting BaFe$_{2}$As$_{2}$ compounds, these experimental features can be adequately reproduced by a scattering model that describes short-ranged and anisotropic spin correlations with overdamped dynamics.
\end{abstract}

\maketitle

\section{Introduction}
The appearance of superconductivity (SC) in doped $A$Fe$_2$As$_2$ materials ($A=$ Ca,Sr,Ba) is linked to the suppression of antiferromagnetic (AFM) ordering found in the parent compounds.\cite{Ni:09,Canfield:09,Chu:09}  Other unconventional superconductors share a similar phase diagram, suggesting that the AFM spin fluctuations may be responsible for pairing of electrons in the SC state.  Certainly, the AFM fluctuation spectrum itself is directly influenced by superconductivity.\cite{Christianson:08n,Lumsden:09,Chi:09} The appearance of a gap and resonance-like feature in the paramagnetic spectrum of these compounds below the superconducting temperature $T_C$ also closely resembles \cite{Uemura:09n}  other unconventional superconductors and highlights the coupling of the iron spins with electronic charge carriers. Recent neutron scattering results elegantly show that the magnetic resonance can infer details about the symmetry of the superconducting gap.\cite{Christianson:08n}

In order to understand the nature of the superconducting pairing, details of the normal state spin fluctuations must be first understood.  These spin fluctuations are expected to be unusual \cite{Yildirim:08,Si:08,Mazin:08,Chubukov:08_2} since magnetic frustration in the tetragonal paramagnetic phase leads to an additional nematic \cite{Chuang:10} degree-of-freedom due to the weak net magnetic coupling of nearest-neighbor Fe sublattices.  The AFM ordering in parent  $A$Fe$_2$As$_2$ compounds below $T_N$ is observed to occur either simultaneously with or after a structural transition from tetragonal to orthorhombic at $T_S$.

The magnetic excitations in  $A$Fe$_2$As$_2$ in stripe AFM ordered parent compounds have been extensively studied \cite{Zhao:08,Mcqueeney:08,Matan:08,Diallo:09,Zhao:09a} and indicate itinerant spin waves whose excitation spectrum can be adequately described by large nearest ($J_1$) and next-nearest-neighbor ($J_2$) in-plane exchange constants, a substantial interlayer coupling ($J_c$), and strong Landau damping $\gamma$.

The primary role of doping is to suppress both the AFM and structural transitions, resulting in strong AFM spin fluctuations. In the context of establishing the connection between spin fluctuations and superconductivity, it becomes imperative to examine the evolution of these spin excitations across the entire phase diagram of doped  $A$Fe$_2$As$_2$. There have been several theoretical and experimental studies on the effects of charge doping on the spin excitations in the superconducting compounds.\cite{Korshunov:08,Yaresko:09,Harriger:09,Pratt:09,Christianson:09co,Inosov:09} In general, these studies show that the normal state spin fluctuations are quasi-two-dimensional (2D) and strongly damped while remaining peaked at ${\bf Q}_{\mathrm{AFM}}$, the magnetic ordering wavevector of the stripe AFM state. It is unclear if the quasi-2D nature and strong damping are features that appear only with sufficient doping, or whether these are common features of the spin fluctuations in the tetragonal phase.

\begin{figure}
\includegraphics[width=0.85\linewidth]{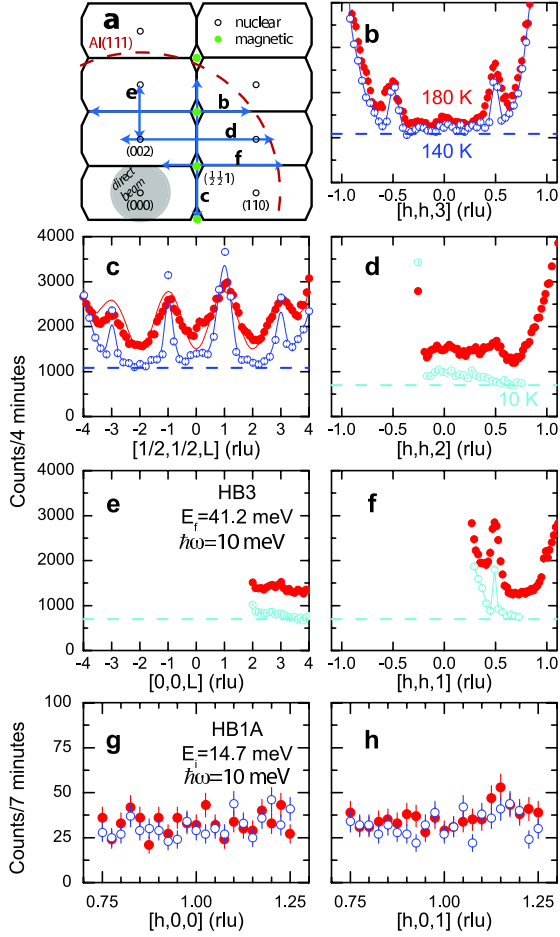}
\caption{\footnotesize Summary of scans performed at $T=$ 10 K, 140 K (empty symbols) and 180 K (filled symbols) on HB3 ($E_{f}=$ 41.2 meV) and HB-1A with spectrometer configurations described in the text. (a) shows $(hhL)$ plane in reciprocal space where the scans at HB3 were performed. (b)-(f) shows the various cuts investigated, as indicated in (a). (g) and (h) show the scans performed in the $(h0L)$ plane using HB1A. No diffuse magnetic scattering was observed in the $(h0L)$ plane. The magnetic signal is only observed centered at wavevectors $\mathbf{Q} = \mathbf{Q}_{\mathrm{AFM}}$.}
\label{Fig1}
\end{figure}

In order to address these questions, we examine the spin correlations that occur in the parent compound  CaFe$_2$As$_2$  above the AFM ordering temperature $T_N$.  Inelastic neutron scattering data on CaFe$_2$As$_2$ single-crystals show that above the simultaneous first order transitions at $T_N = T_S=$ 172 K,\cite{Ni:08} the spin gap collapses and spin wave scattering centered at the stripe ordering wavevector ${\mathbf Q}_{\mathrm{AFM}}$ is replaced by short-ranged and quasi-elastic AFM correlations that extend up to at least 60 meV.  Just above $T_N$ (at $T=180$ K) the low energy magnetic response is quasi-elastic with anisotropic in-plane correlations. We find the in-plane correlation length to be $\xi_{T_{+}} \simeq$ 8 {\AA} along the orthorhombic $a$-axis and $\xi_{T_{-}} \simeq $6 {\AA} along $b$. Weak modulations of the scattered intensity are also observed along the $c$-axis indicating a two-dimensional character to the paramagnetic fluctuations. In general, spin correlations weaken and broaden further in momentum and energy with increasing temperature, but are still observed up to the highest measured temperature of 300 K.
 These observations can be explained in the context of spin dynamics overdamped by particle-hole excitations. In particular, we use a phenomenological theoretical model with in-plane and inter-plane magnetic anisotropy to consistently fit our data for all temperatures, obtaining the ratios $J_{1}/J_2 \simeq 0.55$ and $J_{c}/J_2 \simeq 0.1$.
 We find that the spin fluctuations in the paramagnetic phase of the parent compound bear a close resemblance to the paramagnetic fluctuations in the superconducting compositions.

This article is laid out as follows. In section II below, the experimental conditions under which the experiment were performed and the sample details are presented. The data analysis and results are presented in section III. Finally, a discussion and a summary are given in section IV.


\section{Experimental procedures}

\begin{figure}
\includegraphics[width=1\linewidth]{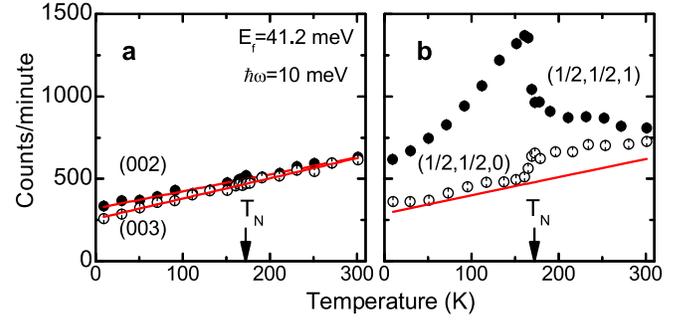}%
\caption{\footnotesize Temperature evolution of the neutron scattering signal measured on HB3 at $\hbar\omega=10$ meV. (a) Background estimate measured away from $\mathbf{Q}_{\mathrm{AFM}}$ at $(002)$ (solid symbols)  and $(003)$ (open symbols) showing no anomaly at $T_{N}$.  Solid lines are linear fits to the temperature dependent intensity. (b) Intensity at $\mathbf{Q}_{\mathrm{AFM}}=(1/2,~1/2,~1)$ and $(1/2,~1/2,~0)$.  The solid line is the non-magnetic background estimate obtained from averaging the fits at $(002)$ and $(003)$, shown in panel (a).}%
\label{Fig2}%
\end{figure}

	Inelastic neutron scattering measurements were performed on a single crystal mosaic ($\sim$400 small single-crystal samples) of CaFe$_{2}$As$_{2}$ with a total mass of $\sim$ 2 grams that are co-aligned to within 1.5 degrees full-width-at-half-maximum (FWHM). The preparation methods of the single-crystals have been described elsewhere. \cite{Ni:08} Data were collected using the HB3 and HB1A triple-axis spectrometers at the High Flux Isotope Reactor at Oak Ridge National Laboratory and the MAPS chopper spectrometer at the ISIS facility at Rutherford Appleton Laboratory. HB3 was operated in relaxed resolution for measurement of the diffuse scattering signals in the paramagnetic phase, with fixed final energy ($E_{f}$) configurations, $E_{f}=$ 14.7 meV and 41.2 meV, and 48'-60'-80'-120' collimation.  The sample was mounted in a closed-cycle refrigerator and oriented for scattering in the tetragonal $(hhL)$ plane. HB1A was operated with fixed incident neutron energy of 14.7 meV and 48'-40'-40'-136' collimation and the sample mounted in the $(h0L)$ plane. The MAPS experiment was performed at $T=$180 K, with an incident energy of 100 meV using the same sample aligned with the $c$-axis along the incident beam direction.

 To avoid confusion, the data is exclusively presented in tetragonal units and we define $\mathbf{Q}=\frac{2\pi}{a}(h\mathbf{i}+k\mathbf{j})+\frac{2\pi}{c} L\mathbf{k}$ as the momentum transfer indexed according to the $I4/mmm$ tetragonal cell with lattice parameters $a$ = 3.88 {\AA} and $c$ = 11.74 {\AA} at 300 K. The vectors $\mathbf{i}$, $\mathbf{j}$ and $\mathbf{k}$ are the fundamental translation unit vectors in real space. For comparison with the AFM low temperature orthorhombic ($o$) structure, we note the following relations between the Miller indices of the two phases, $h$=$(H_o+K_o)/{2}$, $k$=$(H_o-K_o)/{2}$, and $L$=$L_o$. For convenience, we sometimes use the reduced momentum transfer ${\mathbf q}$=${\mathbf Q}-{\mathbf Q}_{\mathrm{AFM}}$ in reciprocal lattice units (rlu) where ${\mathbf Q}$ is the momentum transfer to the sample and ${\mathbf Q}_{\mathrm{AFM}}$=$(h_0,k_0,l_0)$ the reciprocal lattice vector which defines the AFM low temperature zone center. Typical AFM wave
 vectors studied are (1/2,~1/2,~$L$) with $L=\mathrm{odd}$.

After suitable subtractions of the non-magnetic (background) scattering, the observed magnetic inelastic neutron scattering data is cast in terms of the dynamical structure factor $S({\mathbf Q},\omega)$, which is related to the imaginary part of the dynamic spin susceptibility $\chi''({\mathbf Q},\omega)$ via the fluctuation-dissipation theorem,
\begin{equation}
S({\mathbf Q},\omega)=2{(r_0)}^2\frac{{F^2(\mathbf{Q})}}{4\pi \mu_B^2}\frac{\chi''({\mathbf Q},\omega)}{1-e^{-{\hbar\omega/kT}}}
\label{eq_sqw}
\end{equation}
where ${(r_0)}^2=290.6$ mbarns Sr$^{-1}$ is a conversion factor to bring the intensity into absolute units of mbarns meV$^{-1}$Sr$^{-1}$ f.u.$^{-1}$ (Sr=Steradian, f.u.=Formula Unit) and $F(\mathbf{Q})$ is the magnetic form factor for the Fe$^{2+}$ ion.

\begin{figure}
\includegraphics[width=1\linewidth]{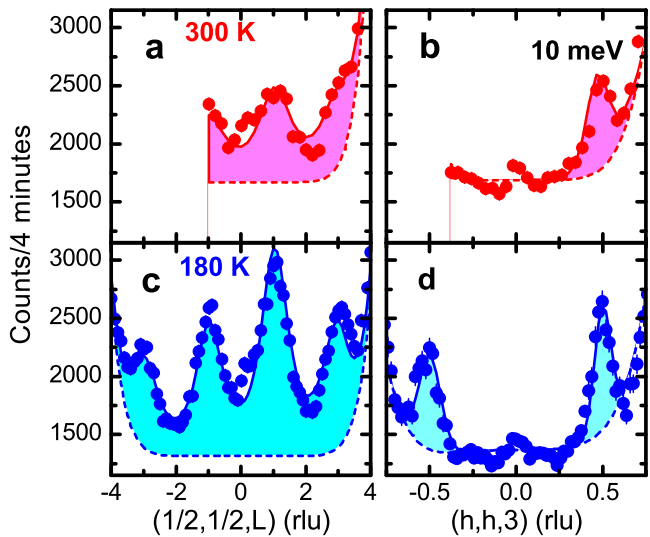}%
\caption{\footnotesize (Color online). $L$ and $h$-dependence of the scattering at $\hbar\omega=$ 10 meV measured on the HB3 instrument with $E_f=$41.2 meV. The scans are performed along the $(1/2,~1/2,~L)$ and $(h,~h,~3)$ directions for temperatures $T=$300 K [(a) and (b)] and $T=$180 K [(c) and (d)]. The solid lines in (a) and (c) correspond to fits to the dynamical susceptibility described in Eq. (\ref{eq_xi}). The solid lines in (b) and (d) are guide to the eye, and based on Lorentzian fits to the data. The dashed line is an estimate of background scattering.  Panels (c) and (d) show the sharper magnetic scattering at 180 K.}%
\label{Fig4}%
\end{figure}

\section{Analysis and Results}

\subsection{Survey of reciprocal space}
Fig. \ref{Fig1} shows the $(hhL)$ plane in reciprocal space where the low energy measurements were performed. It also shows several $Q$-scans taken at temperatures below (10 K and 140 K) and above $T_N$  (180 K)  the concomitant structural and N{\'e}el ordering temperature $T_N=T_S=$ 172 K. Below $T_N$, magnetic Bragg peaks appear at $\mathbf{Q}_{\mathrm{AFM}}=$ $(1/2,~1/2,~L)$ positions with $L=\mathrm{odd}$ that describe the ordered AFM stripe structure. Figs. \ref{Fig1}(b)-(f) show various cuts, as indicated in Fig. \ref{Fig1}(a), through the $(hhL)$ scattering plane at a finite energy transfer of 10 meV and $E_{f}=$ 41.2 meV.  Similar scans performed with $E_f=14.7$ meV show qualitatively the same results. Below $T_N$, sharp excitations are observed at $\mathbf{Q}_{\mathrm{AFM}}$ that originate from very steep spin wave excitations in the ordered state.\cite{Mcqueeney:08,Zhao:08} The difference in sharpness of the spin wave peaks in the $[h,h,0]$ and $[0,0,L]$ directions is due to the anisotropy in the spin wave velocity, as discussed in Ref.[\onlinecite{Mcqueeney:08}]. When the sample is warmed up above $T_N$, strong intensity remains at $\mathbf{Q}_{\mathrm{AFM}}$ position with much broader lineshapes indicating short-ranged AFM spin correlations. The scans shown in Fig. \ref{Fig1} are strongly influenced at higher angles by the presence of aluminum phonon scattering from the sample holder and low angle scattering from the direct beam. Both contributions lead to very high background levels and limit the range of the $Q$-scans.

Since the stripe AFM ordering may be frustrated in the tetragonal structure, we searched for evidence of spin correlations at other wavevectors in addition to the strong components of the diffuse excitations near $\mathbf{Q}_{\mathrm{AFM}}$.    No magnetic diffuse scattering was observed along various symmetry directions in the $(h0L)$ plane.  In particular, no evidence of magnetic scattering was seen at the wavevector (1,0,$L=\text{even}$) corresponding to N{\'e}el (C-type) AFM fluctuations (see Figs. \ref{Fig1}(g) and (h)).  In the $(hhL)$ plane, there are indications of weak peaks in the extended $Q$-scans at wavevectors other than $\mathbf{Q}_{\mathrm{AFM}}$ which may arise from additional magnetic modulations in the paramagnetic phase. For example, very weak peaks can be observed at (002) and (003) (see Figs. \ref{Fig1}(b), (d), and (e)) which would correspond to the presence of ferromagnetic correlations and A-type magnetic correlations (ferromagnetic within the layer, AFM between layers), respectively.

In order to understand the development of correlations at $\mathbf{Q}_{\mathrm{AFM}}$ and address the potential existence of additional magnetic modulations (002) and (003), the temperature dependence was measured at various points in the $(hhL)$ plane at an energy transfer of 10 meV. Fig. \ref{Fig2}(a) shows the temperature evolution of the scattered intensities at (002) and (003). The intensities show no anomaly at $T_N$, but rather the intensity increases linearly with temperature as expected for a phonon background. This background is observed throughout the scattering plane and partly arises from aluminum phonon scattering from the sample holder, with little or no magnetic contribution. We can use the intensity at (002) and (003) as an estimate for this phonon background. In Fig. \ref{Fig2}(b), the temperature dependence at $\mathbf{Q}_{\mathrm{AFM}}=(1/2,~1/2,~1)$ and also $(1/2,~1/2,~0)$ is shown. The intensity at $\mathbf{Q}_{\mathrm{AFM}}$ increases from low temperatures as expected for the increasing Bose population factor of the low lying spin wave modes. At $T_N$, there is a sharp drop in the intensity consistent with the first-order transition to the paramagnetic state. The paramagnetic intensity decreases above $T_N$ and is nearly at background by 300 K. At $(1/2,~1/2,~0)$ the intensity is nearly at background level below $T_N$, increases sharply at the transition, and decreases slowly at higher temperatures.

 \begin{figure}
\includegraphics[width=0.85\linewidth]{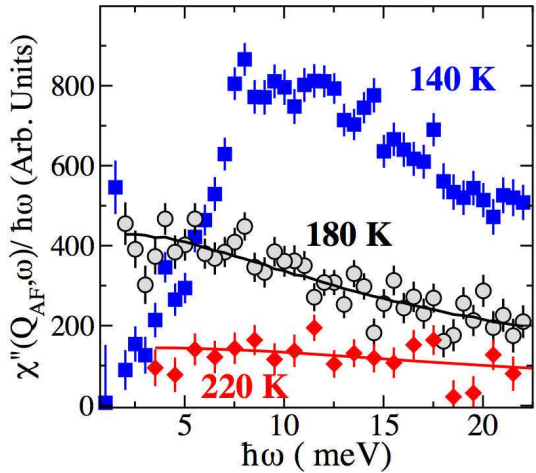}
\caption{\footnotesize (Color online) The dynamic magnetic susceptibility of CaFe$_2$As$_2$ as a function of energy at ${\mathbf Q}_{\mathrm{AFM}} = (1/2,~1/2,~3)$ for $T=140$ K (blue solid squares), $T=180$ K (grey solid circles) and $T=220$ K (red solid diamonds). Above the magneto-structural transition $T_N=T_S=172$ K, a broad magnetic spectrum is observed as quasi-elastic response near ${\mathbf Q}_{\mathrm{AFM}}$. Data taken at $T=180$ K were fit to  a Lorentzian form given in Eq. (\ref{QENS}) convoluted with the instrumental resolution  (black solid line). The Lorentzian half-width $\Gamma_T$ at $T=180$ K is 10 meV. At $T=220$ K, we estimate the energy linewidth to be $\sim13$ meV using the expression $\Gamma_T=\gamma(\frac{a}{\xi_{T}})^2$ and the fitted value of $\gamma$ (temperature independent Landau damping defined in the text) and that of the correlation length $\xi_{T}$ at 220 K. The calculated Lorentzian scattering at $T=220$ K is shown as a red solid line. In contrast, sharp
spin waves having an energy gap $\Delta$ of $\sim$7 meV are observed in the ordered phase at $T=$ 140 K.}
\label{fig_Efits}
\end{figure}

The $L-$dependence of the paramagnetic scattering is consistent with weak antiferromagnetic correlations between layers atop a constant magnetic background, as illustrated in Fig. \ref{Fig2}(b).  Figs. \ref{Fig4}(a) and (c) show the $L$-dependence of the scattering at $\hbar\omega=$ 10 meV and along the $(1/2,~1/2,~L)$ direction for $T=$ 300 K and 180 K, respectively. The $L$-dependence displays sinusoidal variation with maxima at odd values of $L$. The lines shown correspond to fits to the dynamic susceptibility and will be described in detail below. The intensity drops by a factor of 2 between 180 K and 300 K indicating the gradual evolution of the system to less correlated quasi-2D spin fluctuations similar to the reduction in intensity for cuts along the $[h,h,3]$ direction shown in Figs. \ref{Fig4} (b) and (d). The temperature dependence of Al phonons is primarily responsible for the increase in background between 180 K and 300 K. However, the constant magnetic background itself is also weakly temperature dependent as inferred from Fig. \ref{Fig2} (a). We return to this in section \ref{t-dep}.

\begin{figure}
\includegraphics[width=1\linewidth]{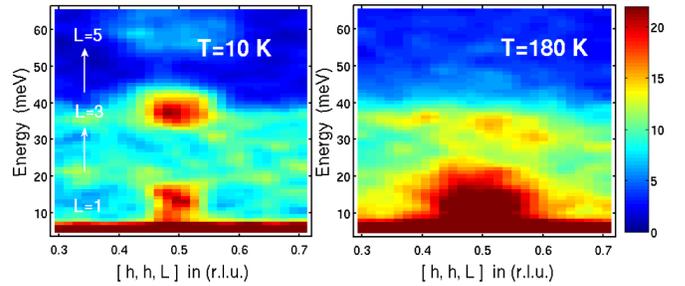}
\caption{\footnotesize (Color online). Magnetic excitations in CaFe$_2$As$_2$ measured on the MAPS spectrometer with an incident energy of $E_i=100$ meV at $T=10$ K (left panel) and 180 K (right panel). The data shows magnetic intensity as a function of the $[h,~h]$ direction and the energy transfer after averaging over the transverse $[h,-h]$ direction in the range 0.4$<h<$0.6. Given the fixed crystal orientation with incident beam along $L$, the $L$ component of the wave vector varies with the energy transfer as indicated.  Excitations below $T_N$ are consistent with steep spin waves and diffuse magnetic excitations are observed above $T_{N}$ at $T=180$ K.}
\label{fig_contoursE}
\end{figure}

In all, the surveys of magnetic scattering intensities above $T_N$ in the $(hhL)$ and $(h0L)$ planes indicate that the AFM spin correlations are restricted to the vicinity of $\mathbf{Q}_{\mathrm{AFM}}$, the wavevector of the stripe ordered phase.  One essential difference in the paramagnetic phase is an increase in the $c$-axis anisotropy and tendency towards 2D spin fluctuations, as indicated by the weakly modulated rod of scattering along $L$.  This is entirely analogous to the behavior of AFM spin fluctuations in the optimally doped superconductors, where interlayer correlations are very weak\cite{Lumsden:09,Inosov:09}.

\begin{figure}
\includegraphics[width=1\linewidth]{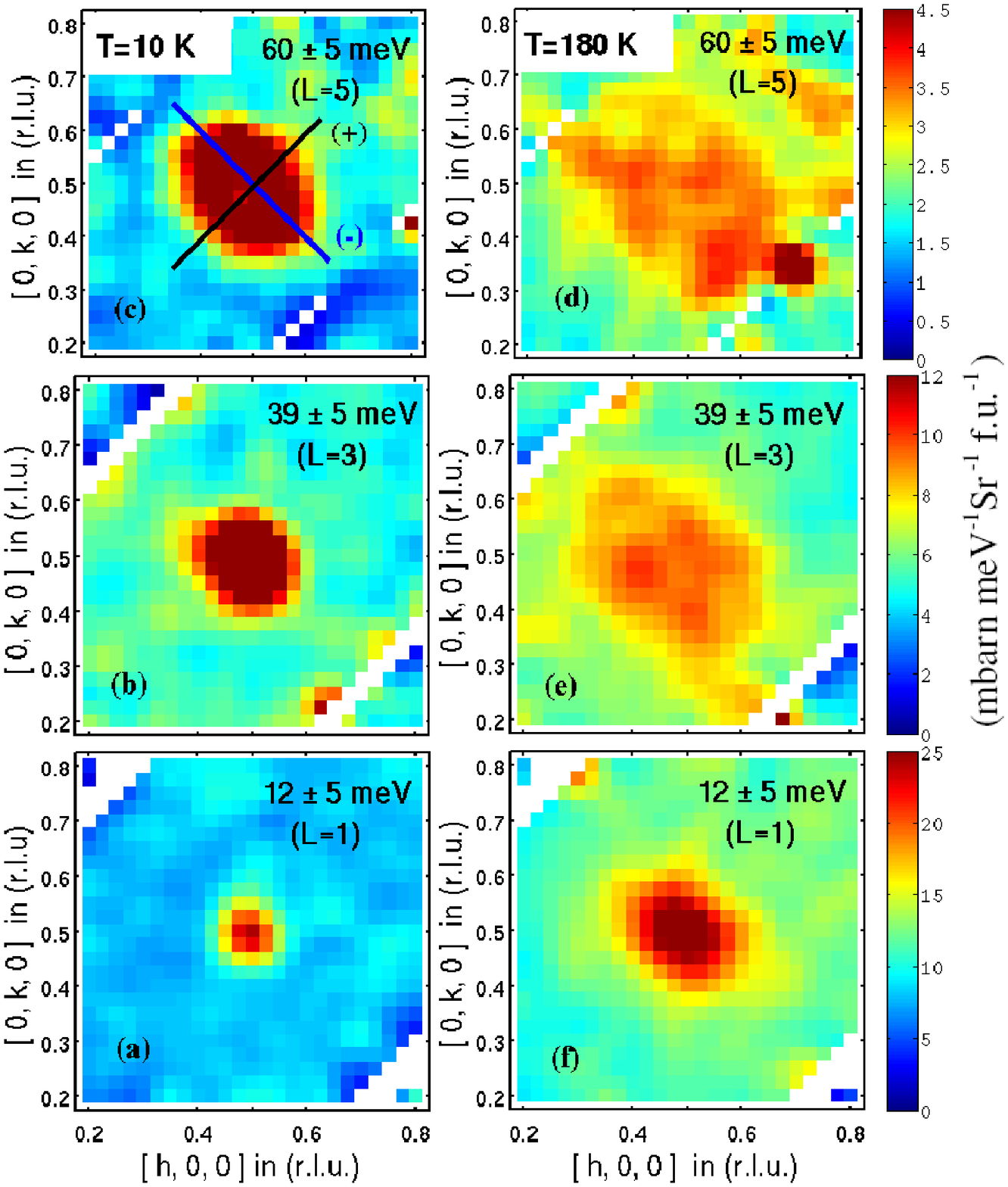}
\caption{\footnotesize (Color online). Constant energy slices ($\Delta E=\pm 5$ meV) through the excitation spectrum of CaFe$_{2}$As$_{2}$ in the $(h,k)$ plane at $T=10$ K and 180 K, as observed on MAPS. Energy slices are chosen to correspond to odd values of $L$.  Intensity shown is in absolute units (mbarn Sr$^{-1}$ meV$^{-1}$ per formula unit). Below $T_N$, well defined spin waves are observed around $\mathbf{Q}_{\mathrm{AFM}}$ (see also Ref. [\onlinecite{Diallo:09}]). Above $T_N$, strong but short-range magnetic correlations remain around ${\mathbf Q}_{\mathrm{AFM}}$ and extend up to at least 60 meV.  Solid lines in panel (c) show the directions along which the cuts in Fig. \ref{fig_E_MAPS} are taken, with (+) designating longitudinal cuts and (-) transverse cuts.}
\label{fig_contours}
\end{figure}

\begin{figure}
\includegraphics[width=0.85\linewidth]{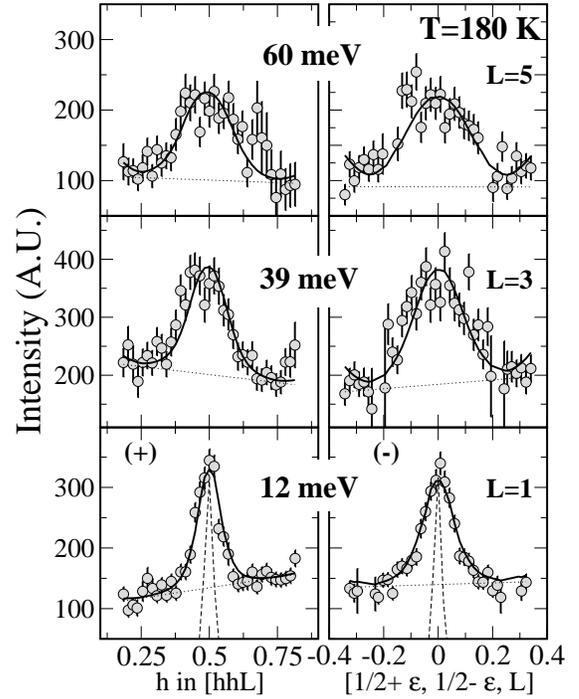}
\caption{\footnotesize  Longitudinal (+) and transverse (-) constant-energy cuts measured around ${\mathbf Q}={\mathbf Q}_{\mathrm{AFM}}$=$(1/2,~1/2,~L)$ on the MAPS spectrometer. The cuts are taken at $\hbar\omega=$12, 39 and 60 meV, which correspond to $L=1, 3$ and 5, respectively. Solid lines are best fits of Eq. (\ref{eq_Inos}) to the data. Dotted lines indicate the fitted background, and the dashed lines show the instrumental resolution  function.}
\label{fig_E_MAPS}
\end{figure}

\subsection{Energy dependence}
Fig. \ref{fig_Efits} depicts the dynamical structure factor $\chi''(\mathbf{Q}_{\mathrm{AFM}},\omega)/\hbar\omega$ at $\mathbf{Q}_{\mathrm{AFM}}=(1/2,~1/2,~3)$ and $\hbar\omega<$ 22 meV for temperatures above and below $T_N$.  In order to obtain $\chi''$ from the raw data, a non-magnetic background was estimated by measurements at $\mathbf{Q}=(0.35,~0.35,~3)$ and subtracted. Below $T_N$, the low energy magnetic spectrum consists of spin waves with a sizeable spin gap of 7 meV (see Ref.~\onlinecite{Mcqueeney:08}). Just above $T_N$, the gap collapses and the spin wave scattering is replaced by gapless, diffusive excitations. At 180 K, the diffusive excitations can be fit to a quasi-elastic Lorentzian form,
\begin{equation}
\frac{\chi''({\mathbf Q}_{\mathrm{AFM}},\omega)}{\hbar\omega}=\frac{A}{(\hbar\omega)^2+\Gamma_T^2}
\label{QENS}
\end{equation}
\noindent with an energy linewidth  of $\Gamma_T=10\pm1$  meV.  The parameter $A$ is an arbitrary intensity scale factor. As the temperature is increased, the Lorentzian half-width grows rapidly. At $T=220$ K, the spectrum weakens considerably with temperature and appears flat within the energy range measured, thus the Lorentzian widths become large and poorly defined. This rapid increase in the quasi-elastic linewidth with temperature is explained below.

\subsection{Spectrum of paramagnetic spin fluctuations near $\mathbf{Q}_{\mathrm{AFM}}$ }\label{t-dep}

At temperatures just above $T_N$ ($T=$ 180 K, $T/T_N=$ 1.05) we used the MAPS spectrometer to perform a detailed survey of the spin fluctuations in the paramagnetic phase in the vicinity of $\mathbf{Q}_{\mathrm{AFM}}$.  For detailed modeling, the MAPS measurements were normalized in absolute scattering units of mb Sr$^{-1}$ meV$^{-1}$ f.u.$^{-1}$ by comparison to a vanadium standard.  Data were collected at 180 K and also in the AFM ordered phase at 10 K (with an incident energy of 100 meV). As in previous work, \cite{Diallo:09}  we use the MSLICE program \cite{Coldea:04} to visualize the data and to take one and two dimensional cuts through main crystallographic symmetry directions for subsequent data analysis with the TOBYFIT suite of analysis programs described below. \cite{Tobyfit:04} Where possible, symmetry equivalent cuts and slices were added to improve statistics. Fig. \ref{fig_contoursE} shows slices of the neutron intensity along the $[h,~h]$ direction as a function of energy transfer after averaging over the $[h,-h]$ direction.  Below $T_N$, the data in the left panel of Fig. \ref{fig_contoursE} show a steep plume of intensity centered at $\mathbf{Q}_{\mathrm{AFM}}$ arising from AFM spin waves.  The sizeable exchange coupling along $c$ leads to variations in the structure factor along $L$, observed as energy-dependent intensity oscillations that are peaked at the AFM zone centers; $\hbar\omega$ = 12 meV ($L$=1), 39 meV ($L$=3), and 60 meV ($L$=5). Analysis of the AFM spin wave spectra is described in detail in Ref. [\onlinecite{Diallo:09}].  Above $T_N$, the right panel of Fig. \ref{fig_contoursE} indicates that the magnetic spectrum is much broader in ${\mathbf Q}$, and energy dependent oscillations are much less pronounced, confirming short-ranged spin correlations within the Fe layer and a weakening of interlayer correlations.

Fig. \ref{fig_contours} shows the neutron intensity for several constant energy slices at $T=$ 10 K and 180 K. The energies are chosen such that $L$ = 1, 3, and 5, in order to measure the spin fluctuations in $(h,~k)$-planes containing $\mathbf{Q}_{\mathrm{AFM}}$ (i.e. measuring the spin correlations within the Fe layers). Below $T_N$, Figs. \ref{fig_contours} (a)-(c) show three  constant energy slices through the AFM spin wave cone centered $\mathbf{Q}_{\mathrm{AFM}}$. Above $T_N$, Figs. \ref{fig_contours} (d)-(f) again demonstrate that the sharp spin waves are replaced by very broad scattering in the paramagnetic phase.  The paramagnetic fluctuations persist up to at least 60 meV and are notably anisotropic with an elliptical shape.  This anisotropic scattering is characterized by transverse $[h,-h]$ cuts through $\mathbf{Q}_{\mathrm{AFM}}$ being broader than the longitudinal $[h,~h]$ cuts, as illustrated in Fig. \ref{fig_E_MAPS} where the solid lines are the resolution convoluted fits of Eq. (\ref{eq_xi}) to the experimental data, described below. This anisotropy implies that the paramagnetic fluctuations are described by two different spin correlation lengths within the Fe layer. An in-plane anisotropy of this type is allowed in the I4/mmm tetragonal cell, i.e. 4-fold symmetry has not been broken. The origin of this anisotropy is discussed below.
\begin{figure}
\includegraphics[width=1\linewidth]{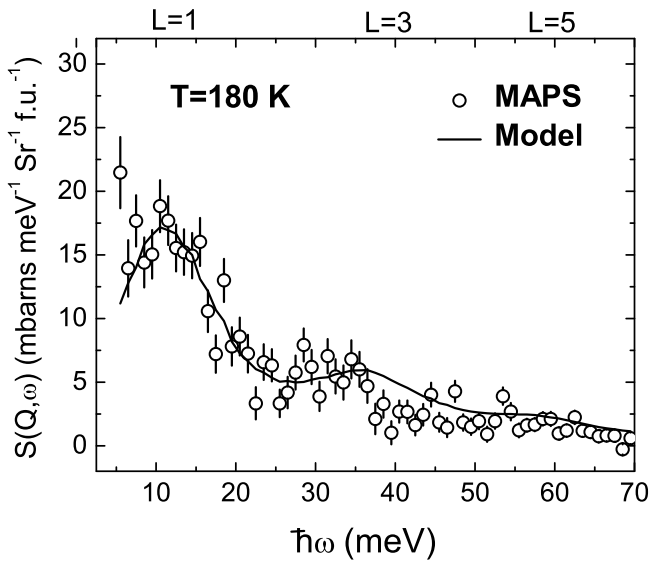}
\caption{\footnotesize Net magnetic neutron scattering intensity as a function of energy measured on MAPS around $\mathbf{Q}_{\mathrm{AFM}}=(1/2,~ 1/2,~ L)$ after subtraction of a non-magnetic background at $\mathbf{Q}=(0.15,~0.5,~L)$. The integration ranges used in MSLICE are such that $\Delta h=\pm0.05$. Lines are the best fits of Eq. (\ref{eq_xi}) to the data.}
\label{fig_MAPS_W}
\end{figure}

 \begin{figure}
\includegraphics[width=0.85\linewidth]{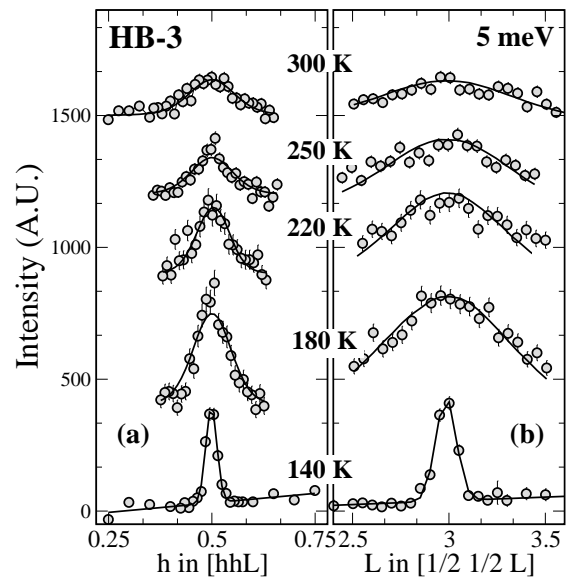}
\caption{\footnotesize Fits of Eq. (\ref{eq_xi}) to constant energy cuts at 5 meV along $h$ and $L$, as a function of increasing temperature ($180\le T\le300$ K). Circles are experimental data measured on HB3 with $E_{f}=$ 14.7 meV. Solid lines are the best fit to the data. Both the $h$- and $L$ scans were fit to Eq. (\ref{eq_xi}) (as shown in Fig. \ref{Fig4} over a wider range of $h$- and $L$). The data at $T=140$ K was fit to a spin wave model (as described in Ref. [\onlinecite{Mcqueeney:08}]), and  is shown here for comparison. }
\label{fig_fits}
\end{figure}

Fig. \ref{fig_MAPS_W} shows the energy dependence of the MAPS neutron intensity at ${\mathbf Q}_{\mathrm{AFM}}=(1/2,~1/2,~ L)$ at $T=180$ K after subtraction of background estimated at $\mathbf{Q}=(0.15,~0.5,~L)$.  In the geometry of our time-of-flight experiment, $L$ varies with energy transfer, however, $L$-dependent oscillations here and in Fig. \ref{fig_contoursE} are only weakly present. After correcting for the dependence of the magnetic form factor, the data was fit to the full susceptibility function described in Eq. \ref{eq_xi} after convolution with the instrumental resolution function.  The model describes the MAPS data quite well, including weak L-dependent oscillations, and gives relaxational half-width $\Gamma_{T} =$ 10$\pm$1 meV, which is consistent with the HB3 measurements shown in Fig. \ref{fig_Efits}.

 Fig. \ref{fig_fits} shows HB3 scans along the $[h,h]$ and $L$-directions through $\mathbf{Q}_{\mathrm{AFM}}=(1/2,~1/2,~3)$ and $\hbar\omega=5$ meV at several temperatures.  Similar to Fig. \ref{Fig1}, the cuts below $T_N$ show sharp spin wave scattering.   The measurements at several temperatures above $T_N$ indicate a gradual weakening of the scattering at $\mathbf{Q}_{\mathrm{AFM}}$ as the temperature is raised, reminiscent of Figs. \ref{Fig4} and \ref{fig_Efits}. As noted above, the magnetic scattering along the $[h,h]$ direction through $\mathbf{Q}_{\mathrm{AFM}}$ is much sharper, as shown in Fig. \ref{Fig4} (d), indicating much longer ranged spin correlations within the Fe layer than between layers.  We see now that the weakening originates from temperature-dependent broadening of the scattering in reciprocal space, indicating a gradual reduction in the spin correlation length with temperature. Similar to Fig. \ref{Fig4}, the $L-$dependence of scans at ${\mathbf Q}_{\mathrm{AFM}}=$(1/2,~1/2,~3) and $\hbar\omega=5$ meV were fit to the susceptibility in Eq. (\ref{eq_xi}) for several temperatures up to 300 K and are shown as solid lines in  Fig. \ref{fig_fits}(b).  The description of this fitting and the interpretation of the results is now described.

\subsection{Theoretical Model}
In this section, we describe the theoretical model used to fit the experimental data. At temperatures below $T_N$, the imaginary part of the generalized susceptibility $\chi''(\mathbf{Q},\omega)$ for  CaFe$_{2}$As$_{2}$ is well described by damped spin wave models.\cite{Mcqueeney:08,Diallo:09,Zhao:09a} Above $T_N$, our neutron scattering data reveal several features that have allowed us to develop a fairly detailed model of the paramagnetic excitations.  First, the excitations are diffusive in character, at least for the measured energy scales ($< 60$ meV).  Second, the excitations are nearly two-dimensional, with  a weak modulation of the scattered intensities along $L$.  Finally, broad scattering is observed within the Fe layers that is anisotropically distributed around $\mathbf{Q}_{\mathrm{AFM}}$.  We are thus forced to define two in-plane spin correlation lengths in order to fully describe the quasi-2D spin fluctuations.

Considering these facts,  we start by writing the total dynamical magnetic susceptibility per formula unit as that of a two-dimensional metallic AFM in the paramagnetic phase,\cite{Moriya,Inosov:09} which we have extended to include the effects of an in-plane anisotropy of the magnetic interactions.

\begin{widetext}
\begin{equation}
\chi_{2D}^{-1}(\mathbf{Q}^{2D}_{AFM}+\mathbf{q}_{||},\omega) = \chi_{0}^{-1}\left[(q_{||}^{2}+\eta q_xq_y)a^2 + (\xi_{T}/a)^{-2}-i\hbar\omega/\gamma \right].
\label{eq_xi1}
\end{equation}
\end{widetext}

\noindent The parameter $\xi_{T}$ defines the magnetic correlation length at temperature $T$ in units of {\AA}, $\chi_{0}$ is the staggered susceptibility, and $\gamma$ denotes the damping coefficient originating from the spin decay into particle-hole excitations.  The dimensionless parameter $\eta$ is used to represent the anisotropy of the in-plane correlation lengths.  The vector $\mathbf{q}_{||}=(q_x,q_y)=\frac{2\pi}{a}(h,k)$ is a 2D vector defined relative to $\mathbf{Q}_{\mathrm{AFM}}^{2D} = \frac{2\pi}{a}(1/2,~1/2)$ and $a=\sqrt{2}a_0$ is the lattice constant of the tetragonal cell with $a_0$ the nearest neighbor Fe-Fe distance.  We can define two correlation lengths in directions longitudinal ($q_{x}=q_{y}$ denoted (+) scan) to $\mathbf{Q}_{\mathrm{AFM}}^{2D}$ with a value $\xi_{T_+}=\xi_{T}(1+\frac{\eta}{2})^{1/2}$ and transverse ($q_{x}=-q_{y}$ denoted (-) scan) to $\mathbf{Q}_{\mathrm{AFM}}^{2D}$ with value ${\xi_{T_-}}=\xi_{T}(1-\frac{\eta}{2})^{1/2}$. Thus, we must have $0\leq\eta<2$ in order for these correlation lengths to be meaningful, which we found indeed to be the case.

The effect of weak interlayer AFM coupling $J_c$ on total dynamic susceptibility can be written as

\begin{equation}
\chi^{-1}(\mathbf{Q}_{\mathrm{AFM}}+\mathbf{q},\omega)=\chi_{2D}^{-1}(\mathbf{Q}^{2D}_{AFM}+\mathbf{q}_{||},\omega) + 2J_{c}\text{sin}^{2}(q_{z} c/4)
\label{eq_3}
\end{equation}

\noindent where the scattering vector is defined as $\mathbf{Q}=\mathbf{Q}_{\mathrm{AFM}}+\mathbf{q}_{||}+q_{z}\mathbf{k}$.  This expression can be motivated using the $1/N$ approach for the action of the collective magnetic degrees of freedom employed in Refs. \cite{Fang:08,Fernandes:10}.  Taking the imaginary part of Eq. (\ref{eq_3}), we obtain the expression

\begin{widetext}
\begin{equation}
\chi''(\mathbf{Q},\omega)=\frac{\hbar\omega\gamma\chi_{0}}{(\hbar\omega)^2+\gamma^2\left[(q^2+\eta q_xq_y)a^2 + (\frac{\xi_{T}}{a})^{-2} + \eta_{c}(1+\text{cos}(\pi L))\right]^2},
\label{eq_xi}
\end{equation}
\end{widetext}

\noindent which we used to fit our inelastic neutron scattering data. Here, the parameter $\eta_{c}=J_{c}\chi_{0}$ controls the strength of the interlayer spin correlations and we have substituted $q_{z}=2\pi(L-1)/c$.  A physical motivation for this expression and an interpretation of these parameters in the context of the iron arsenides is given later in section IV.

\begin{table}
\caption{\footnotesize Best fit parameters of Eq. (\ref{eq_xi}) obtained self consistently from both HB-3 and MAPS at temperature $T=$180 K. The parameter $\chi_{0}$ was solely obtained from the MAPS data which were normalized in absolute units. }
\begin{ruledtabular}
\begin{tabular}{| c | c | c |  c | c |}
\hline
$\gamma$ &   $\xi_{T}$ & $\chi_{0}$ & $\eta$ & ${\eta_c}$ \\
 (meV) &  $({\AA}$) & $({\mu_B}^2$ meV$^{-1}$ f.u.$^{-1})$ & & \\
\hline
43$\pm$ 5 & 7.9$\pm$0.10 & 0.2$\pm$0.05  & 0.55$\pm$0.36 & 0.20$\pm$0.02  \\
\hline
\hline
\end{tabular}
\end{ruledtabular}
\label{tbl1}
\end{table}

As we are only able to define $\eta$ from a single MAPS data set at 180 K, we make the assumption that the parameters $\chi_{0}$, $\gamma$, $J_c$ and $\eta$ in Eq. (\ref{eq_xi}) do not depend on temperature, making $\xi_{T}$ the only temperature dependent parameter in the model. However, one can introduce effective temperature dependent parameters $X_{T}=\chi_{0}(\frac{\xi_{T}}{a})^{2}$ and $\Gamma_{T}=\gamma(\frac{a}{\xi_{T}})^{2}$.  With these definitions, the dynamical susceptibility along the longitudinal direction at,  $\mathbf{q}=\left(\frac{q}{\sqrt{2}},\frac{q}{\sqrt{2}},L=1\right)$, or in the 2D limit where $J_{c}=0$ can be written as,

\begin{equation}
\chi_{2D}''(\mathbf{Q}^{2D}_{AFM}+\mathbf{q},\omega)=\frac{\hbar\omega\Gamma_{T} X_{T}}{(\hbar\omega)^2+\Gamma_T^2(1+q^2\xi_{T_+}^2)^2}
\label{eq_Inos}
\end{equation}

At $q=0$, this model reduces to the quasi-elastic Lorentzian lineshape used to describe the susceptibility in Eq. (\ref{QENS}).

In the context of the iron arsenides, Eq. (\ref{eq_Inos}) was first used successfully by Inosov {\em et al.} \cite{Inosov:09} to describe the Co-doped BaFe$_2$As$_2$ superconducting compound. In their work, Inosov {\em et al.} \cite{Inosov:09} assumed a mean-field behavior for the correlation length $\xi_{T}\propto\left(T-T_{N}\right)^{1/2}$. Here, given the strong first-order character of the magnetic transition in the parent compound CaFe$_{2}$As$_{2}$, we do not assume any particular form for temperature dependence of $\xi_{T}$.

For detailed comparison of model susceptibility in Eq. (\ref{eq_xi}) to the data, we use standard inelastic neutron scattering computer programs that account for resolution effects.  Specifically, the high energy data collected on MAPS was analyzed using TOBYFIT \cite{Tobyfit:04} while the low energy response measured on HB3 (up to 15 meV) was analyzed using RESLIB. \cite{Zheludev}

Our protocol was to establish the value of the temperature independent parameters using data sets from both HB3 and MAPS at 180 K where scattering features are more sharply defined.  Data fits at subsequent temperatures would then only require the refinement of a single parameter, $\xi_{T}$.  Due to the interrelated nature of the fitting parameters, self-consistency checks were performed for model fittings on the HB3 and MAPS data sets.  The in-plane anisotropy parameter $\eta$ was  obtained independently from the MAPS data by fitting the longitudinal (+) and transverse (-) cuts, as depicted in Fig. \ref{fig_MAPS_W}.  The observed in-plane anisotropy is then determined from these fitted widths according to $\eta=2{\frac{\xi_{T_+}^2-\xi_{T_-}^2}{\xi_{T_+}^2+\xi_{T_-}^2}}$. The temperature independent damping parameter, $\gamma$, was obtained from both the HB3 and MAPS energy scans at $T=180$ K where we have the best signal to noise ratio. The corresponding fits are illustrated in Figs. \ref{fig_Efits} and  \ref{fig_E_MAPS}. The fits were performed by constraining the value of $\gamma=\Gamma_T(\frac{\xi_{T}}{a})^{2}$ such that the Lorentzian width $\Gamma_T$ remains fixed at 10 meV (fitting only for $\xi_{T}$). Once $\gamma$ was determined at $T=180$ K, it was held fixed in all subsequent fits.

To determine $\eta_c$ at 180 K, we have performed a chi-squared analysis of fits in which only $\eta_c$  ($0\le \eta_c \le 0.5$) was allowed to vary, and other parameters held fixed. The best fits were obtained for $0.9<\eta_c<0.11$, as indicated in Fig. (\ref{Fig4}), which also corresponds to a minimum  $\chi^2$ value. Outside this range, $\chi^2$ drastically increases and the fits are altered. The parameter $\eta_c$ in Eq. (\ref{eq_xi}) which describes the interlayer coupling was found to be $\simeq$ 10\% at $T=180$ K.

\begin{table}
\caption{Temperature evolution of the correlation length $\xi_{T}$ (in {\AA}) for several incident energies, as measured on HB-3. For a given temperature, $\xi_{T}$ appears largely independent of energy, as expected. The final column, containing the correlation length averaged over all energies, is the value used in subsequent model calculations.}
\begin{ruledtabular}
\begin{tabular}{| c | c | c | c | c |}
\hline
        &5 meV & 10 meV  &  15 meV  & $<\xi_{T}>$ ({\AA})  \\
  \hline
  180 K & 8.1$\pm$0.2 & 7.9$\pm$0.1 & 7.8$\pm$0.3 & 7.9$\pm$0.1    \\
\hline
 220 K & 7.0$\pm$0.2 & 7.2$\pm$0.3 & 6.9$\pm$0.2 & 7.0$\pm$0.1 \\
\hline
250 K & 6.0$\pm$0.1  & 6.0$\pm$0.2 & 6.2$\pm$0.2 & 6.1$\pm$0.1 \\
\hline
 300 K & 4.9$\pm$0.1 & 4.9$\pm$0.2 & 4.9$\pm$0.2 & 4.9$\pm$0.1 \\
\hline
\hline
\end{tabular}
\end{ruledtabular}
\label{tbl2}
\end{table}

The parameters $\chi_0$ and $\xi_{T}$ were found to be strongly correlated.  The prescription was used to fix $\xi_{T}$ (effectively fixing the width) and then vary the scale factor $\chi_0$ to get the best fit.  This procedure was performed for constant energy cuts at several energy transfers on both HB3 and MAPS data sets at $T=$ 180 K.  Upon convergence to reasonable values of $\chi_0$ and $\xi_{T}$, the whole procedure was repeated for consistency. Once the best fitted value of $\chi_0$ was obtained at 180 K, we kept it fixed at that same value for all other temperatures and fit the data by varying only $\xi_{T}$.

The MAPS data was normalized to vanadium standard, allowing us to report $\chi_0$ in absolute units. In contrast, the HB3 data was not in absolute units. Thus, the corresponding best fitted scale factor from HB3 gives a value that is proportional to the true $\chi_0$. Therefore the $\chi_0$ value reported here are obtained solely from the MAPS data. In all, we arrived at the following final set of parameters at $T=$ 180 K; $\gamma=43\pm5$ meV,  $\chi_{0}=0.20\pm0.05$ $\mu_B^2$ meV$^{-1}$f.u.$^{-1}$, $\xi_{T_+}=8\pm1$ {\AA} and $\xi_{T_-}=6\pm1.5$ {\AA}. Fits are shown as solid lines in Figs. \ref{Fig4}, \ref{fig_Efits}, \ref{fig_E_MAPS} and \ref{fig_MAPS_W}. From these values, we estimate the average in-plane anisotropy parameter $\langle\eta\rangle\simeq$ 0.55$\pm$0.36, and thus kept $\eta$ fixed at this value in subsequent model fits to the higher temperature data. The corresponding value of the static correlation length $\xi_{T}$ obtained from MAPS is $7\pm2$ {\AA}, in good agreement with the low energy HB3 value of 7.9$\pm$0.1 {\AA}.  A summary of all of the model parameters at $T=$ 180 K are listed in Table \ref{tbl1}.

 Fig. \ref{fig_fits}(a) shows the temperature dependence of the HB3 constant energy cuts along the $[1,1,0]$ direction at 5 meV for several temperatures up to 300 K.  These data were fit to determine the values of $\xi_{T}$ at different temperatures. The correlation length decreases by a factor of $\sim$2 from 180 K to 300 K. Similar fits were performed at constant energy cuts of 10 meV and 15 meV (not shown), giving comparable results. The results are summarized in Table \ref{tbl2}.

We note that the set of parameters in Tables \ref{tbl1} and \ref{tbl2} describe the full data set at all temperatures and energy transfers exceptionally well, as can be seen by reviewing all of the fitted curves in Figs. \ref{Fig4}, \ref{fig_Efits}, \ref{fig_E_MAPS}, \ref{fig_MAPS_W} and \ref{fig_fits}.
For example, the line through the data in Fig. \ref{fig_Efits} at $T=$ 220 K is not a fit, rather it is a model calculation based on the fixed parameters in Tables \ref{tbl1} and \ref{tbl2}.  This universal agreement with the data at all measured energies and temperatures is a strong endorsement for the validity of the nearly AFM model.


\section{Discussion and Conclusions}
\subsection{Discussion}
Despite the strong first order transition and the possible influence of magnetic frustration in the tetragonal phase, we find that the spin fluctuations observed above the AFM ordering temperature in CaFe$_2$As$_2$ are peaked at $\mathbf{Q}_{\mathrm{AFM}}$, the wavevector of the stripe AFM structure. We found no evidence of substantial magnetic fluctuations at any of the other wavevectors studied, indicating that the relevant spin fluctuations throughout the phase diagram are short-ranged fluctuations of the low-temperature stripe AFM ordering. In addition, we find weaker interlayer coupling immediately above $T_N$ in the parent compound, suggesting that the quasi-2D spin fluctuations are a property of the tetragonal paramagnetic phase in general, and does not necessarily occur only after doping.

The substantial spin fluctuations that persist above $T_N$ should have an effect on the bulk transport and magnetic properties of CaFe$_{2}$As$_{2}$.  For example, such spin fluctuations are a likely source of the high resistivity found for CaFe$_{2}$As$_{2}$ between $T_N$ and 300 K at ambient pressure.\cite{Ni:08}  Large spin disorder scattering at ambient pressure is supported by dramatic decrease of the resistivity \cite{Torikachvili:08,Yu:09} and disappearance of spin fluctuations \cite{Pratt:09b} upon transition to the collapsed tetragonal phase under applied pressure.

\begin{table}
\caption{The derived staggered susceptibility $X_{T}$, relaxational linewidth $\Gamma_T$, and bulk susceptibility $\chi_{\mathrm{bulk}}$, as a function of temperature from $180<T<300$ K.}
\begin{ruledtabular}
\begin{tabular}{| c | c | c | c | c |}
\hline
 Temp.        & 180 K & 220 K & 250 K & 300 K \\
\hline
$X_{T}$ {\footnotesize ($\mu_B^2$ meV$^{-1}$ f.u.$^{-1}$)}&0.84 & 0.66 & 0.49 & 0.32\\
$\Gamma_{T}$ {\footnotesize (meV)}&10 & 13 & 17 & 26\\
$\chi_{\text{bulk}}$ {\footnotesize (10$^{-4}$ emu mol$^{-1}$)}&2.41 & 2.40 & 2.39 & 2.37\\
\hline
\end{tabular}
\end{ruledtabular}
\label{tbl3}
\end{table}

The bulk magnetic susceptibility of CaFe$_{2}$As$_{2}$ also displays a weak, non Curie-Weiss temperature dependence above $T_N$.\cite{Ni:08} Even though Eq. (\ref{eq_xi}) is strictly valid only as an expansion for momenta close the AFM ordering vector, we can use this expression for the dynamic spin susceptibility to estimate the $\mathbf{Q}=0$ bulk susceptibility as $\chi_{\text{bulk}}=\chi(\mathbf{Q}=0,\omega=0)=\chi_{0}/[2\pi^{2}(1+\eta/2)+ (\frac{\xi_{T}}{a})^{-2}+2\eta_{c}]$.

For large correlation lengths, the bulk susceptibility takes on a temperature independent value of $\chi_{\text{bulk}}\approx \chi_{0}/[1+2\pi^2(1+\eta/2)+2\eta_{c}]=$ 2.43x10$^{-4}$ emu mol$^{-1}$ f.u.$^{-1}$. This estimate has the same order of magnitude as the bulk susceptibility determined from magnetization measurements, \cite{Ni:08} further justifying the validity of the model.  The shorter correlation lengths found well above $T_N$ will decrease the bulk susceptibility as shown in Table \ref{tbl3}, although this effect is small for the temperatures studied.

The extrapolation of Eq.~(\ref{eq_xi}) to other momentum values also allows us to estimate the size of the fluctuating moment per iron atom

\begin{equation}
\langle m^2\rangle=\frac{1}{2}\frac{3\hbar}{\pi}\frac{\int{ \chi''({\mathbf Q},\omega){(1-e^{-{\hbar\omega/kT}})}^{-1}{d\mathbf Q}d\omega}}{\int{ d\mathbf Q}}.
\label{eq_m2}
\end{equation}
  
The $Q$-integration is performed in the orthorhombic zone (containing one ellipse of scattering) defined by the ranges $0 \leq Q_{x}(Q_{y}) \leq \frac{2\pi}{a}$ and $0 \leq Q_{z} \leq \frac{4\pi}{c}$.  The factor of {1/2} in eq. \ref{eq_m2} converts the results from squared-moment per formula unit to squared moment per iron.  If we choose a high energy cut-off of $\sim$200 meV corresponding to the upper limit of observed spin wave excitations,\cite{Diallo:09,Zhao:09a} we obtain $\sqrt{\langle m^2 \rangle}\simeq$ 0.7 $\mu_B$ per iron which is very close to the size of the observed ordered moment. \cite{Goldman:08}

The spin fluctutations observed above the AFM ordering temperature in CaFe$_2$As$_2$ bear a close resemblance to that of doped superconducting BaFe$_2$As$_2$.  Based on the model fitting, the similarity of the spin fluctuations in the parent and superconducting compositions can be compared more quantitatively.  The paper by Inosov {\em et al.} \cite{Inosov:09} describes the parameters for short-range spin correlations in the optimally doped superconducting composition using an equation similar to (\ref{eq_Inos}).  We can estimate the values of the model parameters at 180 K for this SC composition using mean-field temperature scaling as described by Inosov {\em et al}.  Also, since the model parameters described in their work assume isotropic in-plane spin correlations, we extended their model by including the anisotropy parameter, $\eta$, obtained from our results (i.e. assuming $\eta$ to be both temperature and composition independent).  With these assumptions, we estimated the values of the model parameters for the superconducting sample with the following results at 180 K; the static susceptibility at the antiferromagnetic wavevector is $X_{T}$ = 0.18 $\mu_B^2$ meV$^{-1}$ f.u.$^{-1}$, the longitudinal correlation length is $\xi_{T_+}$ = 11 {\AA}, and the Landau damping $\gamma$ = 180 meV.  This can be compared to the values in CaFe$_2$As$_2$ at the same temperature as found here; $X_{T}$ = 0.9 $\mu_B^2$ meV$^{-1}$ f.u.$^{-1}$, $\xi_{T_+}=$ 8 {\AA}, and $\gamma$ = 43 meV.  The results show that the Landau damping parameter is much larger in the superconducting composition, whereas the correlation lengths at $180$K are similar.  In comparing these numbers, it is important to also recall the differences between the two materials investigated. For instance, in the sample studied by Inosov {\em et al.},\cite{Inosov:09} the alkaline earth metal is Ba and no magnetic phase transition is present. On the other hand, in the sample analyzed here, the alkaline earth metal is Ca and, at $180$K, the system is very close to a strong first-order magnetic transition.

We now turn to a discussion of the physical interpretation of the data.  The new feature observed in our results is the anisotropy of the in-plane spin fluctuations in the paramagnetic phase, represented by the parameter $\eta$.  As discussed above, Eq. (\ref{eq_xi}) is based on a metallic AFM model and a proper starting point to understand this anisotropy is an electronic band model for the magnetic susceptibility. The unique band structure of the iron arsenides, consisting of circular hole pockets at $\mathbf{Q}$=(0,~0) and elliptical electron pockets at $\mathbf{Q}_{\mathrm{AFM}}^{2D}$=(1/2,~1/2), combined with electron-electron interactions will lead to an anisotropic spin response, even in the tetragonal phase.\cite{Graser:09,Zhang:10} More physical insight can be gained by considering a phenomenological model for the collective magnetic degrees of freedom of the pnictides. \cite{Fang:08,Xu:08,Fernandes:10} In these materials, the magnetic lattice in each iron arsenide plane can be subdivided in two interpenetrating antiferromagnetic sublattices, each containing one iron per unit cell. These two sublattices are only loosely coupled, because the local magnetic field produced by one sublattice on the other vanishes in the completely ordered state. To dominant order, the only couplings between the two sublattices that are allowed by symmetry are the biquadratic term $-g(\mathbf{m}_{1}\cdot\mathbf{m}_{2})^{2}$ and the momentum-dependent quadratic term $\eta q_{x}q_{y}\mathbf{m}_{1}\cdot\mathbf{m}_{2}$. Evaluation of the spin-spin correlation function using this two-sublattice model leads to Eq. (\ref{eq_xi}). Within this description, the anisotropy of the correlation length results from an inter-sublattice coupling.

This phenomenological model can be motivated by either an itinerant \cite{Eremin:09} or a localized moment picture \cite{Si:08}.  Indeed, the short correlation lengths encountered in the paramagnetic phase suggest a minimum description where only the short-range interactions are kept. One then has the so-called frustrated $J_{1}$-$J_{2}$ model, with antiferromagnetic nearest-neighbor exchange interaction $J_{1}$ and next-nearest neighbor exchange $J_{2}$. The anisotropy parameter $\eta$ becomes simply the ratio between the two exchange constants, $\eta=J_{1}/J_{2}$, and $J_{2}$ is related to the bare susceptibility $\chi_{0}$ of Eq. (\ref{eq_xi}) through $\chi_{0}=2/J_{2}$.  This also allows us to equate the interlayer coupling parameter $\eta_{c}$ with the ratio $2J_{c}/J_{2}$.

The exchange ratios in the paramagnetic phase can be compared directly to the values in the orthorhombic AFM ordered phase as obtained from measurements of spin waves \cite{Diallo:09,Zhao:09a} (see also Fig. \ref{fig_contours}) From the observed anisotropy of the in-plane correlation lengths, we can determine $J_{1}/J_{2} = 0.55 \pm 0.36$ in the tetragonal phase, implying that $J_1$ is indeed antiferromagnetic.  In the AFM ordered orthorhombic phase, there are distinct nearest-neighbor exchange interactions, $J_{1a}$ and $J_{1b}$ and the experimentally determined ratio \cite{Diallo:09,Zhao:09a} $(J_{1a}+J_{1b})/2J_{2}\approx 0.6 - 1$ is similar in magnitude to the exchange ratio in the paramagnetic phase and within the error bars. Thus, the relative strength of the average exchange interactions in the Fe plane do not appear to be strongly affected by the structural transition. The value of this ratio $J_{1}/J_{2} < 2$ places the paramagnetic phase of CaFe$_2$As$_2$ in a regime of frustrated magnetism within the $J_{1}-J_{2}$ model and consistent with the presence of spin nematic correlations.\cite{Fang:08,Xu:08,Fernandes:10}  Taking the fitted value of $\chi_{0}$ and considering the Fe moment to be $\sim1\mu_{B}$, one can use the relation $\chi_{0}\sim 2/J_{2}$ to estimate the the effective next-nearest neighbor exchange $J_{2}\approx$ 10 meV, which can be compared with the value of 25-35 meV obtained from fits to the spin wave dispersion in the ordered phase.

With regard to the $2D$ nature of the spin fluctuations above $T_{N}$, we can estimate the relative strength of the interlayer coupling in the paramagnetic phase through the parameter $\eta_c$.  The resulting exchange ratio $J_{c}/J_{2} = 0.1 \pm 0.01$, is lower than the ratio $J_{c}/J_{2}\approx 0.15 - 0.25$ in the ordered AFM state\cite{Zhao:09a}, implying that the fluctuations have a more two-dimensional character above $T_N$. This result is consistent with angle-resolved photoemission spectroscopy (ARPES),\cite{Chang:09} which showed that the band structure dispersion of CaFe$_{2}$As$_{2}$ is quasi-2D above $T_N$ but 3D below $T_N$.

\begin{figure}
\includegraphics[width=0.8\linewidth]{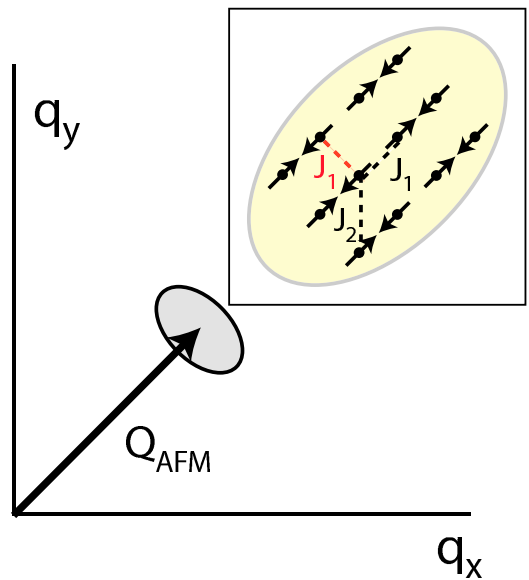}
\caption{\footnotesize Schematic drawing of the anisotropic scattering in the Fe layer at $\mathbf{Q}_{\mathrm{AFM}}$.  The inset shows roughly the extent of spin correlations in the layer which result in anisotropic scattering at $\mathbf{Q}_{\mathrm{AFM}}$.  Correlations transverse to $\mathbf{Q}_{\mathrm{AFM}}$ are shorter due to the frustrating effect of antiferromagnetic $J_1$ interactions on ferromagnetic bonds (shown in red), whereas $J_1$ strengthens correlations along the antiferromagnetic bonds extending in the longitudinal direction.}
\label{Fig10}
\end{figure}

At last, we would like to mention that the existence of anisotropic in-plane correlation lengths has a simple physical interpretation within the frustrated $J_{1}-J_{2}$ model. The magnetically ordered phase consists of diagonal stripes with the magnetic moments ordered ferromagnetically along one diagonal and antiferromagnetically along the correspondent orthogonal diagonal, as illustrated in Fig. \ref{Fig10}. The direction longitudinal to $\mathbf{Q}_{\mathrm{AFM}}$ points along chains of antiferromagnetic spins while the transverse direction points along ferromagnetic chains. Since $J_{1}$ and $J_{2}$ are antiferromagnetic couplings, the nearest-neighbor bonds along the transverse direction are frustrated, leading to a shorter correlation length when compared to the correlation length associated to the unfrustrated bonds along the longitudinal direction.  Thus, the data are consistent with nanoscale regions of dynamic AFM stripe/nematic correlations.

\subsection{Summary}
In conclusion, paramagnetic excitations in CaFe$_{2}$As$_{2}$ ($T_{S}=T_{N}=172$ K) have been investigated in the temperature range from $180$ K ($\sim 1.05T_N$) up to 300 K ($1.8T_{N}$).  Above $T_N=$ 172 K, the spin gap collapses and antiferromagnetic spin wave scattering is replaced by in-plane anisotropic short-ranged AFM correlations that extend to high energy ($\hbar\omega<$60 meV) and temperature ($T<$ 300 K). The paramagnetic excitations are observed only around the low temperature magnetic zone center $\mathbf{Q}_{\mathrm{AFM}}$. The in-plane anisotropy of the spin correlation length has a natural explanation in terms of short-ranged magnetic interactions in the $J_1-J_2$ model.  Thus, the paramagnetic fluctuations correspond to the emergent stripe AFM ordered structure and are suggestive of nematic correlations.  The magnetic correlations above $T_N$ have an anisotropic 2D character and bear a close resemblance to the paramagnetic fluctuations observed in the superconducting compositions.

\section{Acknowledgments}
We thank V.P. Antropov, D. Johnston and J. Schmalian for stimulating discussions. Work at the Ames Laboratory was supported by the Department of Energy, Basic Energy Sciences under Contract No. DE-AC02-07CH11358. Technical assistance of the staff at ORNL and ISIS is gratefully acknowledged.


\end{document}